\documentclass[superscriptaddress,secnumarabic,nobibnotes,aps,prd,showkeys,showpacs,nofootinbib,preprint]{revtex4}

\usepackage{graphicx}
\usepackage{float}
\usepackage{bm}
\usepackage{amsmath}
\usepackage{amsfonts}
\usepackage{amssymb}
\usepackage{epstopdf}
\usepackage{natbib}%
\newcommand{\bee}{\begin{equation}}
\newcommand{\eee}{\end{equation}}
\newcommand{\eaa}{\end{eqnarray}}
\newcommand{\baa}{\begin{eqnarray}}
\def\ni{\noindent}

\usepackage{color}

\begin{document}

\title{Non-Gaussian thermostatistical considerations \\ upon the Saha Equation}

\author{Br\'aulio B. Soares}\email{brauliosoares@uern.br}
\affiliation{Departamento de Ci\^encia e Tecnologia, Universidade do Estado do Rio Grande do Norte, Natal, RN, Brazil}
\author{Ed\'esio M. Barboza Jr.}\email{edesiobarboza@uern.br}
\affiliation{Departamento de F\'isica, Universidade do Estado do Rio Grande do Norte, 59610-210, Mossor\'o, RN, Brazil}
\author{Everton M. C. Abreu}\email{evertonabreu@ufrrj.br}
\affiliation{Departamento de F\'{i}sica, Universidade Federal Rural do Rio de Janeiro, 23890-971, Serop\'edica, RJ, Brazil}
\affiliation{Departamento de F\'{i}sica, Universidade Federal de Juiz de Fora, 36036-330, Juiz de Fora, MG, Brazil}
\affiliation{Programa de P\'os-Gradua\c{c}\~ao Interdisciplinar em F\'isica Aplicada, Instituto de F\'{i}sica, Universidade Federal do Rio de Janeiro, 21941-972, Rio de Janeiro, RJ, Brazil}
\author{Jorge Ananias Neto}\email{jorge@fisica.ufjf.br}
\affiliation{Departamento de F\'{i}sica, Universidade Federal de Juiz de Fora, 36036-330, Juiz de Fora, MG, Brazil}

%%%%%%%%%%%%%%%%%%%%%%%%%%%%%%%%%%%%%%%%%%%%%%%%%%%%%%%%%%%%%%%%%%%%%%%%%%%%%%%%%%%%%%%%%%%%
% 52.25.kn thermodynamcs of plasmas
% 82.60.-s chemical thermodynamics
% 05.70.-a thermodynamics
%%%%%%%%%%%%%%%%%%%%%%%%%%%%%%%%%%%%%%%%%%%%%%%%%%%%%%%%%%%%%%%%%%%%%%%%%%%%%%%%%%%%%%%%%%
\begin{abstract}
\noindent The Saha equation provides the relation between two consecutive ionization state populations, like the Maxwell-Boltzmann velocity distribution of the atoms in a gas ensemble.   Saha equation can also consider the partitions functions for both states and its main application is in stellar astrophysics population statistics.   
This paper presents two non-Gaussian thermostatistical generalizations for the Saha equation: the first one towards the Tsallis nonextensive $q$-entropy and the other one is based upon Kaniadakis $\kappa$-statistics.  Both thermostatistical formalisms are very successful when used in several complex astrophysical statistical systems and we have demonstrated here that they work also in Saha's ionization distribution.   
%We have analyzed complex atoms and pair production systems.
We have obtained new chemical $q$-potentials and their respective graphical regions with a well defined boundary that separated the two symmetric intervals for the $q$-potentials.  The asymptotic behavior of the $q$-potential was also discussed.   Besides the proton-electron, we have also investigated the complex atoms and pair production ionization reactions.
\end{abstract}
%%%%%%%%%%%%%%%%%%%%%%%%%%%%%%%%%%%%%%%%%%%%%%%%%%%%%%%%%%%%%%%%%%%%%%%%%%%%%%%%%%%%%%%%%%%%
\date{\today}

\maketitle
%%%%%%%%%%%%%%%%%%%%%%%%%%%%%%%%%%%%%%%%%%%%%%%%%%%%%%%%%%%%%%%%%%%%%%%%%%%%%%%%%%%%%%%%%%%%%%

\maketitle

\section{Introduction}

In the current literature there is no doubt concerning the importance of the nonexponential distributions arising from a striking number of complex statistical systems. Besides, the nonextensivity (or nonadditivity) appears to take an increasingly important role when making an analysis of the universal phenomena. Furthermore, the power-law distributions that emerge from nonextensive entropies provide a strong motivation to apply their frameworks to several physical and non-physical problems.
Among these physical problems there is the one about ionization fractions, which provides vital informations to understand several physical systems, as those related to the production of neutrinos in the solar core, among others related to astrophysics and cosmology \cite{pessah01a,pessah01b}.

The Saha equation (also known as Saha-Langmuir equation) is the theoretical mechanism that establishes the ionization fractions expected for a system.
Therefore, the Saha ionization formula \cite{saha20,saha22} has played a very important role in the current progress of astrophysics. As an example, in the calculation of the physical conditions in various solar formation environments, the Saha equation is used frequently. Besides, we can mention that different models for the Saha equation are discussed in the literature (e.g. \cite{milne28a,milne28b,milne28c,chandra30,ecker63,rouse64,stewart66}). It is also important to remark that a seemingly generalization of the Saha equation by means of the $q$-entropy, has been released in \cite{pessah01a,pessah01b}.  

The thermodynamic path for the construction of a modified version of the Saha equation concerning the case of the so-called two-temperature or multi-temperature plasmas is a very popular issue in theoretical works (the interested reader can see Refs. \cite{uns} and references therein).  The main subject involving these works is the different forms of the chemical-equilibrium equation with some controversy questioning which one of those forms is the precise one to use.
This scenario had a certain impact on the literature considering its applications \cite{outros}.  For example, van de Sanden {\it et al} \cite{sanden}, a few decades ago, introduced an interesting thermodynamic generalization of the Saha equation for a two-temperature plasma.  The respective assumption was that the internal
energy states of the heavy particles are ruled by Boltzmann's law where $T = T_e$.  It was also pointed out that the multi-temperature Saha equation (MSE) is not valid.  After that, Bakshi \cite{bakshi} constructed another modified version of the Saha equation for a two-temperature plasma, which has considered the under-population of the excited states thanks to the deviation from local thermodynamic equilibrium.

The purpose of this paper is to introduce formally two different non-Gaussian statistical formulations of the Saha equation concerning basically the Tsallis \cite{Tsallis88} and Kaniadakis \cite{Kaniadakis01a} thermostatistics.    We agree with Suyari \cite{Suyari04} when he says that \emph{these statistics reveal surprising mathematical structures which might reveal an even more surprising physics behind,}  and in fact we have discovered new features.

To organize our paper we will follow a sequence such that in Section II the usual Saha distribution ionization model is described. In Section III 
%we have presented the Tsallis statistics also known as nonextensive (NE) statistics. In Section IV 
we have extended the Saha model by means of Tsallis formalism. In Section IV 
%we explained the Kaniadakis statistics, also known as kappa statistics. In Section VI 
we generalized the Saha equation by using the Kaniadakis framework. The conclusions and final remarks are depicted in Section V.

%Thus, next sections discuss the non conventional statistics involved in the present %work. First we deal with a process of the ionization of hydrogen. Then we propose a %simple generalization for the function of partioning by means of the generalized %statistics from Tsallis and Kaniadakis frameworks.
%############################################################################ 

\section{Ionization reaction and the Saha equation}

Let us start by clarifying some aspects of the ionization reaction considering an hydrogen atom with only one energy level (ground state)

\begin{equation}
\label{reac}
    H^{+} + e^{-} \longleftrightarrow H^{o} + |\epsilon_{o}|\,\,,
\end{equation}

\ni where $|\epsilon_o |=13,6\,eV$ is the ground state binding energy.  Hence, it is natural to await the recombination temperature to rely on the barion to photon ratio together with the ionization energy.   To compute the fractional ionization $y$, which is a function of both $T$ and the baryon to photon ratio, demands some statistical mechanics, which is the objective here.   

Let us assume that the particles in Eq. \eqref{reac} are both in thermal (same temperature) and kinetic equilibrium, which means that the distribution of energy and momentum obeys a Bose-Einstein or Fermi-Dirac distribution.  If we say that the reaction in Eq. \eqref{reac} is in chemical equilibrium, it is tantamount to say that the reaction rate is the same whatever the direction of the reaction (arrow) occurs (points).

When we have that a certain reaction is in statistical equilibrium at a determined temperature $T$, the number $n_a$ of particles having mass $m_a$ can be given by the well known Maxwell-Boltzmann formula
\bee
\label{max-boltz}
n_a\,=\,g_a\bigglb(\frac{m_a k_B T}{2 \pi \hbar^2} \biggrb)^{3/2}\,exp \bigglb(-\frac{m_a c^2}{k_B T} \biggrb)\,\,,
\eee

\ni where  $k_{B}$ the Boltzmann constant, $\hbar$ is the Planck constant, $g_{s}$ is the statistical weight from the particle spin or its relative degeneracy, and $m$ the particle mass.  We also have that $k_B T \ll m_a c^2$ for nonrelativistic particles.   Considering the reaction in Eq. \eqref{reac} involving $H$, $e^-$ and $p$, we can write an equation that connects their number densities
\bee
\label{aaaa}
\frac{n_H}{n_p n_e}\,=\,\frac{g_H}{g_p g_e} \bigglb( \frac{m_H}{m_p m_e} \biggrb)^{3/2}\,\bigglb(\frac{k_B T}{2\pi\hbar^2} \biggrb)^{-3/2}\,\exp \bigglb[\frac{(m_p + m_e - m_H)\,c^2}{k_B T}\biggrb]\,\,,
\eee

\ni which can be simplified since $m_H / m_p = 1$, the binding energy can be given by the sum of masses at the exponential factor, the statistical weights $g_e =g_p = 2$ and $g_H = 4$.  This simplification will be accomplished just below.

Considering $n_{p}$, $n_{e}$ and $n_{H}$ as the protons ($H^{+}$), electrons ($e^{-}$) and neutral hydrogen ($H^{o}$) number density respectively, and with $n_{p}=n_{e}$.  Then  we can define the ionization fraction as 
\begin{equation}
    y\equiv\frac{n_{p}}{n_{p}+n_{H}}=\frac{n_{p}}{n}=\frac{n_{e}}{n},
\end{equation}

\ni where the total number density is given by $n=n_{p}+n_{H}$.
When the reaction in Eq. (\ref{reac}) reaches the equilibrium then the chemical potential of protons ($H^{+}$) and electrons ($e^{-}$) should be equal to the chemical potential of $H^{o}$, or, equivalently $\mu_{p}+\mu_{e}=\mu_{H}$.   It is important to say that we are specifically interested in the chemical potential difference between two possible states.   Considering the reaction in Eq. \eqref{reac}, if $\mu_{H} > \mu_{p}+\mu_{e}$, therefore we have the reaction running preferentially from the higher energy state, i.e., $H^o\,+\,|\epsilon_o |$ to the lower energy state, i.e., $p\,+\,\epsilon^-$.   On the other hand, if  $\mu_{H} < \mu_{p}+\mu_{e}$, the reaction runs favorably in the opposite direction.

Assuming that for all the components in such reaction the chemical potential is related, by default, to the number density and that the temperature is given by the simplified version of Eq. \eqref{aaaa}, we have:
\begin{equation}
\label{cp}
e^{\beta\mu}=\frac{n\hbar^{3}}{g_{s}}\left(\frac{2\pi}{m k_{B}T}\right)^{3/2},
\end{equation}

\ni where $\beta\equiv 1/k_B\,T$. Thus we can write
\begin{equation}
n=\frac{g_{s}}{\hbar^{3}}\left(\frac{mk_{B}T}{2\pi}\right)^{3/2}e^{\beta\mu},
\end{equation}
which leads us to the standard exponential form of the Saha equation
\begin{equation}
\frac{n_{e}n_{p}}{n_{H}}=
\left(\frac{m_{e}k_{B}T}{2\pi\hbar^{2}}\right)^{3/2}e^{-\beta|\epsilon_{o}|} \,\,,
\end{equation}

\ni which relates the ionization fraction, $y=n^+/n = n_e/n$, of a gas in thermal equilibrium  to the temperature and pressure (remember that for an ideal gas $P=N\,k_B\,T$).  This equation is used to the ionization of an atom. Molecules and molecular ions have other degrees of freedom, like vibration and rotation.  Because of that the equation needs to be modified in such a way to take these features into account.

A more precise thermodynamic demonstration of the Saha equation modified in a way to show a two-temperature
plasma system can be constructed relied upon the underlying thermodynamic principle that the entropy of an
isolated system will have its maximum at the equilibrium state. It can be shown that we
have to worry about the chemical potential expressions apropos the
two-temperature plasma, since anything goes wrong, it may cause mistakes in the resulting
two-temperature Saha equation \cite{ch}.   So, considering the partition function
\begin{equation}
\label{8}
\frac{n_{e}n_{i}}{n_{a}}= \frac{2Q_i (T)}{Q_a (T)}\,
\left(\frac{m_{e}k_{B}T}{2\pi\hbar^{2}}\right)^{3/2}e^{-\beta|E_{i}|}.
\end{equation}

\ni where $m_e$ is the electron mass, $T$ is the plasma temperature,  $E_i$ is the effective gas ionization energy, which includes the lowering of the ionization
energy caused by the magnetic and/or electric fields inside the plasma \cite{boulos}, $Q_i$ and $Q_a$ are the internal
partition functions of ions and atoms, respectively.  The partition functions are given by
\baa
\label{9}
Q_a &=& \sum_k g_{a,k} \,e^{-\beta |E_{a,k}|} \nonumber \\
Q_i &=& \sum_i g_{i,k} \,e^{-\beta |E_{i,k}|}
\eaa

\ni where $g_{a,k}$ and $g_{i,k}$ are the degrees of degeneracy,  $E_{i,k}$ or $E_{a,k}$ are the energy difference between the $k$-th energy state
and the ground state of the ions and atoms, respectively.  Moreover, Eq. \eqref{8} is also known as the Saha equation including the partition functions, although, originally his expression did not contain the partition functions accounting for the excited states inside ions and atoms.

%literature since Saha (1920) first gave an expression similar to equation \eqref{8} for the equilibrium constant or the mass action law of the ionization–recombination reaction \eqref{reac}, although

%############################################################################ 

\section{The Tsallis' path to Saha analysis}

Tsallis' statistics \cite{Tsallis88}, which is the nonextensive (NE) extension of the Boltzmann-Gibbs (BG) statistical theory, defines also a nonadditive entropy as
\begin{eqnarray}
\label{nes}
S_q =  k_B \, \frac{1 - \sum_{i=1}^W p_i^q}{q-1}\;\;\;\;\;\;\qquad \Big(\,\sum_{i=1}^W p_i = 1\,\Big)\,\,,
\end{eqnarray}

\ni where $p_i$ is the probability of a system to exist within a microstate, $W$ is the total number of configurations (microstates) and 
$q$, known in the current literature, is the Tsallis parameter or NE  parameter.   It is a real (or not, see \cite{complex}) parameter which measures the degree of nonextensivity. 

The definition of entropy in Tsallis statistics carries the standard properties of positivity, equiprobability, concavity and irreversibility. This approach has been successfully used in many different physical systems. Considering some instances, we can mention the Levy-type anomalous diffusion \cite{levy}, turbulence in a pure-electron plasma \cite{turb} and gravitational and cosmological systems \cite{sys, sa, nos1, nos2, nos3, nos4, moradpour1, SGS}.
It is noteworthy to stress that Tsallis thermostatistics formalism has the BG statistics as a particular case in the limit $ q \rightarrow 1$, where the standard additivity of the entropy can be recovered.  In other words, it is mandatory to obtain the BG statistics when we have $q \rightarrow 1$ in Tsallis formalism equations.

Our proposal here is to consider a $q$-exponential formulation for the fugacity $z$ such that the original form is a particular case of this new function. 
Therefore, we consider that the fugacity is related to the chemical potential $\mu$ for every component of the system by 
\begin{equation}
\label{partq}
z(q;\beta,\mu)\equiv e^{\beta\mu}_{q}=\Lambda\frac{n\hbar^{3}}{g_{s}}\left(\frac{2\pi}{m k_{B}T}\right)^{3/2}\,,
\end{equation}

\ni where $$e^{x}_{q}\equiv[1+(1-q)x]^{\frac{1}{1-q}}\,\,,$$ together with $q$, characterize the Tsallis statistics of the system.  Moreover,  
$\Lambda$ is a proportionality constant which should be defined as one when $q=1$, whichever the involved particle. The $\Lambda$ parameter relates the standard fugacity to the $q$-exponential. %Thus, such a fugacity is to be a multiple of the classical one.
 
Then, generalizing Eq. (\ref{partq}) for a particle $i$ we have that
\begin{equation}\label{eque1}
e^{\beta\mu_{i}}_{q}=\Lambda\frac{n_{i}\hbar^{3}}{g_{s_{i}}}\left(\frac{2\pi}{m_{i}k_{B}T}\right)^{3/2}\,.
\end{equation}

\ni Applying the generalized $q$-logarithm on both sides, we can write that
\begin{equation}\label{muq}
\beta\mu_{i}=\ln_{q}\left[\Lambda\frac{n_{i}\hbar^{3}}{g_{s_{i}}}\left(\frac{2\pi}{m_i k_{B}T}\right)^{3/2}\right]\,,
\end{equation}

\ni where $$\ln_{q}x\equiv(x^{1-q}-1)/(1-q)\,\,.$$  Eq. (\ref{muq}) recovers the standard form of the chemical potential of the electron as 
\begin{equation}
\frac{\mu_{e}}{k_{B}}=-T\,\ln\left(\frac{T^{\frac{3}{2}}}{n_{e}}\right)-\frac{3}{2}T\,\ln\left(\frac{2\pi m_{e}k_{B}}{h^{2}}\right)-T\,\ln 2 \,,
\end{equation}

\ni when $q=1$ and $\Lambda=1$.

From Eq. (\ref{reac}) we can obtain a particular case of the ionization fraction given by
\begin{equation}
\frac{n_{e}n_{p}}{n_{H}}=\frac{D_{(e,p;H)}}{\Lambda}\bigglb[\frac{m_{e}m_{p}k_{B}T}{(m_{p}+m_{e})2\pi\hbar^{2}}\biggrb]^{3/2}\frac{e^{\beta\mu_{e}}_{q}e^{\beta\mu_{p}}_{q}}{e^{\beta(\mu_{H}+|\epsilon_{o}|)}_{q}}\,,
\end{equation}

\ni where \(D_{(e,p;H)}\equiv g_{s_{e}}g_{s_{p}}/g_{s_{H}}\) is obtained from the  degeneracies. The quantity $\beta|\epsilon_{o}|$ in the $q$-exponential arises from the binding energy of the hydrogen atom. Using the ionization fraction definition, \(y\equiv n_{p}/(n_{p}+n_{H})=n_{p}/n=n_{e}/n\) and \(m_{e}\ll m_{p}\longrightarrow m_{H}\approx m_{p}\), we can rewrite the equation above as 
\begin{equation}
\label{eq19}
\frac{y^{2}}{1-y}= \frac{D_{(e,p;H)}}{\Lambda\,n}\left(\frac{m_{e}k_{B}T}{2\pi\hbar^{2}}\right)^{3/2}\frac{e^{\beta\mu_{e}}_{q}e^{\beta\mu_{p}}_{q}}{e^{\beta(\mu_{H}+|\epsilon_{o}|)}_{q}}\,.
\end{equation}

\ni Let us use the $q$-algebraic properties \cite{borges98} to provide  
\begin{equation}\label{qalge2}
\frac{e^{\beta\mu_{e}}_{q}e^{\beta\mu_{p}}_{q}}{e^{\beta(\mu_{H}+|\epsilon_{o}|)}_{q}}\,=\, \exp_q \!\!\left[\frac{\beta\mu_{e}+\beta\mu_{p}+(1-q)\beta^{2}\mu_{e}\mu_{p}-\beta\mu_{H}-\beta|\epsilon_{o}|}{1+(1-q)(\beta\mu_{H}+\beta|\epsilon_{o}|)}\right]\,.
\end{equation}

\ni Classically the system reaches equilibrium whenever $\mu_{e}+\mu_{p}-\mu_{H}=0$. However, we can define a new chemical $q$-potential for balance as 
\begin{equation}
\label{muq2}
\mu_{(q)}\equiv\mu_{e}+\mu_{p}+(1-q)\beta\mu_{e}\mu_{p}-\mu_{H}\,,
\end{equation}
or

\begin{equation}
\label{mq}
\mu_{(q)}\equiv\mu_{cl}+(1-q)\beta\mu_{e}\mu_{p}\,,
\end{equation}
where $\mu_{cl}\equiv\mu_{e}+\mu_{p}-\mu_{H}$.

It is important to notice some features about the new chemical $q$-potential. The first one is that $\mu_{(q)}$, Eq. (\ref{mq}), 
changes more rapidly with $q$ and a decreasing temperature behavior. The second one is that, for a fixed temperature, $\mu_{(q)}$ decreases more rapidly as the $\mu_e\mu_p$ product increases. Besides, whenever $q\rightarrow1$ or $k_B\,T\gg\mu_e\mu_p$ then $\mu_{(q)}\rightarrow\mu_{cl}$. 
On the other hand, whenever $q=1$ then $\mu_{cl}=0$ such that $q=1$ recovers the classical statistics of equilibrium. Therefore, if $q$ represents the statistics of the particles involved in the reaction, $q=1$ represents the equilibrium statistics. Then one can assume that $\mu_{(q)}$ represents the chemical potential in some grand canonical type ensemble of the Tsallis statistic. In Fig. 1 we have plotted the nonextensive chemical $q$-potential, $\mu_{(q)}$, Eq. (\ref{mq}), as a function of the temperature $T$ for different values of $q$.

Notice in Fig. 1 that the behavior of the $q$-potential at $q=1$ is very clear.   Through the straight line at $q=1$, we can see that this line acts like a boundary separating the two $q$-regions, i.e., one for positive and the other for negative values of $q$.   The symmetrical behavior of the curves is a very interesting result.   It shows that, whatever the sign of $q$, the $q$-potential behavior is the same.  Fig. 1 shows also the asymptotic behavior of the $q$-potential as function of both the temperature and $q$-parameter.   This asymptotic behavior is independent of the value or the sign of the $q$-parameter, except for $q=1$, as explained above.

\begin{figure}[H]
	\centering
	\includegraphics[width=8.0cm]{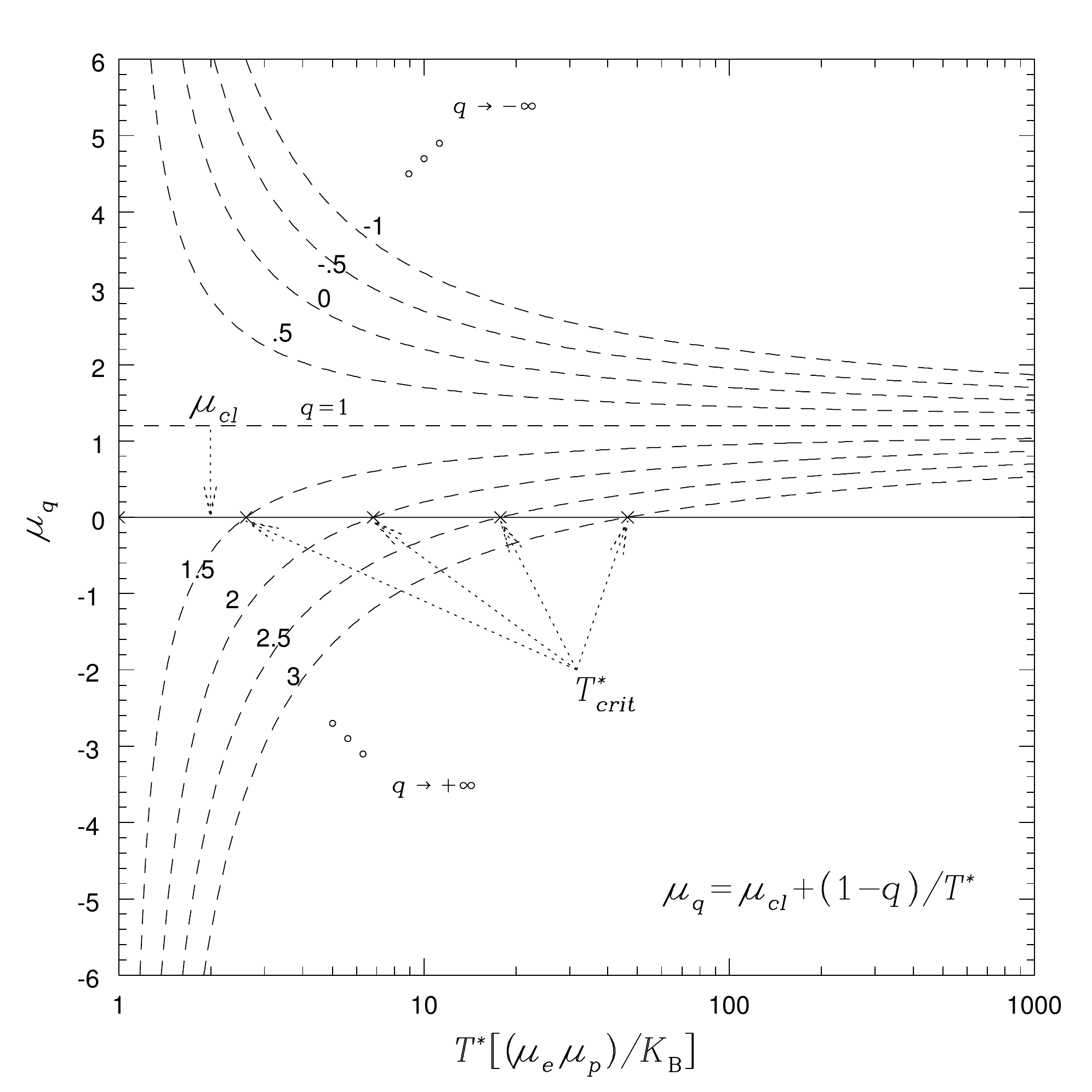}
	\caption{The nonextensive chemical $q$-potential, $\mu_{(q)}$, plotted as a function of the temperature T. %$$T_{crit}^{*}=(q-1)/\mu_{cl}$%}
		\label{means}}
\end{figure}

%\revision{\(\dfrac{n_i}{\eta_i}=\left(\dfrac{m_i\,K_B\,T}{2\pi\,\hbar^2}\right)^{3/2}%\)}

\noindent Moreover, Eqs. (\ref{eq19}), (\ref{qalge2}) and (\ref{muq2}) lead to the ionization fraction or the $q$-generalized Saha equation
 
\begin{equation}\label{qalge4}
 \frac{y^{2}}{1-y}= \frac{1}{\Lambda\,n}\left(\frac{m_{e}k_{B}T}{2\pi\hbar^{2}}\right)^{3/2}
\exp_q \bigglb[\frac{\mu_{(q)}-|\epsilon_{o}|}{k_{B}T+(1-q)(\mu_{H}+|\epsilon_{o}|)}\biggrb]\,,
\end{equation}
or

\begin{equation}\label{qalge5}
\frac{y^{2}}{1-y}= \frac{4.01\times 10^{-9}}{\Lambda}\,\frac{T^{3/2}}{\rho}\,e^{\Theta_{q}}_{q}\,,
\end{equation}
where $\rho$ is the density in grams per cubic centimeter, $T$ is the Kelvin temperature and $\Theta_{q}$ is defined as

\begin{equation}
\label{argtheta}
\Theta_{q}\equiv\left[\frac{\mu_{(q)}-|\epsilon_{o}|}{k_{B}T+(1-q)(\mu_{H}+|\epsilon_{o}|)}\right]\,.
\end{equation}

As an immediate mathematical consequence, Eq. (\ref{qalge2}) is constrained by imposing that we have the following condition $$\beta\mu_{H}+\beta|\epsilon_{o}|\neq\frac{1}{q-1}\,\,.$$ Then for $k_{B}=8.6\times 10^{-5}\,\textrm{eV}\,\textrm{K}^{-1}$ and $|\epsilon_{o}|=13.6\,\textrm{eV}\,\,$ we have that $$\mu_{H}\neq(8.6\times 10^{-5}\,\textrm{eV}\,\textrm{K}^{-1})\frac{T}{q-1}-13.6\,\textrm{eV}\,\,.$$

\ni The factor $\Theta_{q}$ of the $q$-exponential in Eq. (\ref{qalge5}) implies that for $q\neq1$, the ionization fraction, other than temperature and density, is also dependent upon the chemical $q$-potential of the hydrogen released in  reaction (\ref{reac}).   This constraint leads to some assumptions.   For example, the chemical $q$-potential of hydrogen as a function of temperature and density may have a different behavior than it is expected for a classical behavior. If this is not the case, one can expect that the binding energy for the atomic hydrogen is different from $13.6\,eV$. Finally, both cases can occur, even though it is difficult to be verified. It can be demonstrated that
Eq. (\ref{qalge5}) recovers the classical equilibrium equation 
\begin{equation}
  \label{qalge7} \frac{y^2}{1 - y} = 4.01\times 10^{-9}\,\frac{T^{3/2}}{\rho}\,e^{-\beta|\epsilon_0|}\,,
\end{equation}
when $q = 1$ and $\Lambda=1$.

%\section{Other consequences}
\subsection{Complex Atoms}

Consider the following reaction for the $r$-th ionization of the atom $Z$
\begin{equation}
 Z^{r+1}+e\longleftrightarrow Z^{r}+\delta\epsilon\,,
\end{equation}

\ni where $\delta\epsilon$ represents the energies from released photons.   Taking the partition  function in Eq. (\ref{partq}) for the above equation we have

\begin{equation}
 z_i=e^{\beta\mu_i}_{q}=\Lambda\,\frac{n_i\hbar^{3}}{g_{s_i}}\left(\frac{2\pi}{m_i k_{B}T}\right)^{3/2}\,.
\end{equation}

\noindent Analogously as we discussed above, the $r$-th ionization yields 
\begin{equation}
e^{\beta\mu_r}_{q}=\Lambda\,\frac{n_r\hbar^{3}}{g_{s_r}}\left(\frac{2\pi}{m_r k_{B}T}\right)^{3/2}\,,
\end{equation}
together with
\begin{equation}
e^{\beta\mu_{r+1}}_{q}=\Lambda\,\frac{n_{r+1}\hbar^{3}}{g_{s_{r+1}}}\left(\frac{2\pi}{m_{r+1} k_{B}T}\right)^{3/2}\,,
\end{equation}
and
\begin{equation}
e^{\beta\mu_e}_{q}=\Lambda\,\frac{n_e\hbar^{3}}{g_{s_{e}}}\left(\frac{2\pi}{m_e k_{B}T}\right)^{3/2}\,.
\end{equation}

\noindent Hence, we obtain that
\begin{equation}
\frac{n_{r+1}}{n_{r}}n_{e}=\frac{1}{\Lambda}
\frac{g_{s_{r+1}}g_{s_{e}}}{g_{s_{r}}}
\frac{1}{\hbar^3}\left(
\frac{m_{r+1}m_{e}k_{B}T}{m_r2\pi}
\right)^\frac{3}{2}\frac{e^{\beta\mu_{r+1}}_{q}e^{\beta\mu_{e}}_{q}}{e^{\beta(\mu_{r}+\delta\epsilon)}_{q}}\,.
\end{equation}

%As \(m_{r+1}/m_r\approx 1\), and defining \(D_{(e,r+1;r)}\equiv g_{s_{r+1}}g_{s_{e}}/g_{s_{r}}\), in addition to $q$-algebraic properties in equation (\ref{qalge2}) we have
\noindent Similarly to Eq. (\ref{qalge2}) we can write

\begin{equation}
 \frac{e^{\beta\mu_{r+1}}_{q}e^{\beta\mu_{e}}_{q}}{e^{\beta(\mu_{r}+\delta\epsilon)}_{q}}
\,=\,\exp_q \!\!\left[\frac{\beta\mu_{r+1}+\beta\mu_{e}+(1-q)\beta^{2}\mu_{r+1}\mu_{e}-\beta\mu_{r}-\beta\delta\epsilon}{1+(1-q)(\beta\mu_{H}+\beta\delta\epsilon)}\right]\,.
\end{equation}

\noindent Let us define
\begin{equation}
\mu_{(q)}\equiv\mu_{r+1}+\mu_{e}+(1-q)\beta\,\mu_{r+1}\,\mu_{e}-\mu_{r}\,,
\end{equation}

\noindent namely, 
\begin{equation}
\mu_{(q)}\equiv\mu_{cl}+(1-q)\beta\,\mu_{r+1}\,\mu_{e} \,,
\end{equation}

\noindent with $\mu_{cl}\equiv\mu_{r+1}+\mu_{e}-\mu_{r} $ and considering that $ m_{r+1}/m_r\approx 1 $ we can write that

\begin{equation}
\label{33}
 \frac{n_{r+1}}{n_{r}}n_{e}=\frac{D_{(e,r+1;r)}}{\Lambda}\frac{1}{\hbar^3}\left(\frac{m_{e}k_{B}T}{2\pi}\right)^{3/2}
 \,\exp_q \!\!\left[\frac{\mu_{(q)}-\delta\epsilon}{k_{B}T+(1-q)(\mu_{r}+\delta\epsilon)}\right]\,,
\end{equation}

\noindent where the parameter $D_{(e,r+1;r)}$,  a factor of degenerescence, is defined as $ D_{(e,r+1;r)}\equiv g_{s_{r+1}}/g_{s_{r}}$.   Therefore, Eq. \eqref{33} is analogous to the Saha equation for  complex atoms.   Next we will consider another ionization reaction, the pair production.

\subsection{Pair production}

Consider the  electron-positron annihilation reaction

\begin{equation}
e^+ + e^-\longleftrightarrow \gamma\,,
\end{equation}

\noindent where $\gamma$ represents the released photon energy. %and \(\mu_{+}+\mu_{-}=\delta\).% 
Taking the partition function in Eq. (\ref{partq}) for the above equation we can write
\begin{equation}
\label{parti}
 z_{(+,-)}=e^{\beta\mu_{(+,-)}}_{q}=\Lambda\,\frac{n_{(+,-)}\hbar^{3}}{g_{s_{(+,-)}}}\left(\frac{2\pi}{m_{(+,-)} k_{B}T}\right)^{3/2}\,,
\end{equation}

\noindent where the $(+,-)$ signs are relative to $e^+$ and to $e^-$, respectively. Thus, from Eq. (\ref{parti}) we have
\begin{equation}
\label{parti2}
n_+\,n_- =\frac{1}{\Lambda^{2}}\,g_{s_{+}}g_{s_{-}}(m_{+}m_{-})^{3/2}\left(\frac{k_{B}T}{2\pi\hbar^2}\right)^{3}e^{\beta\mu_{+}}_{q}e^{\beta\mu_{-}}_{q}\,.
\end{equation}

\noindent Using the equation

\begin{equation}
\label{re1}
e^{\beta\mu_{+}}_{q}\,e^{\beta\mu_{-}}_{q}=\exp_q \Big[\beta\mu_{+}+\beta\mu_{-}+(1-q)\beta^{2}\mu_{+}\mu_{-}\Big]\,,
\end{equation}

\noindent and the property  that \(\beta\mu_{+}=-\beta\mu_{-}\) in Eq. (\ref{re1}) we have 

\begin{equation}
\label{re2}
e^{\beta\mu_{+}}_{q}\,e^{\beta\mu_{-}}_{q}=e^{-(1-q)\beta^{2}\mu_{-}^{2}}_{q}\,.
\end{equation}

\noindent Then, using Eqs. (\ref{re1}) and (\ref{re2}) into (\ref{parti2}) we have
\begin{equation}
n_+\,n_- =\frac{1}{\Lambda^{2}}\,g_{s_{+}}g_{s_{-}}(m_{+}m_{-})^{3/2}\left(\frac{k_{B}T}{2\pi\hbar^2}\right)^{3}e^{-(1-q)\beta^{2}\mu_{-}^{2}}_{q}\,.
\end{equation}

\noindent As \(m_{+}\,m_{-}=m_e^2\) and, defining a degenerescence factor \(D_{(+,-)}\equiv g_{s_{+}}\,g_{s_{-}}\), we have
\begin{equation}
n_+\,n_- = \frac{D_{(+,-)}}{\Lambda^{2}}\,\left(\frac{m_e k_{B}T}{2\pi\hbar^2}\right)^{3}e^{(q-1)\beta^{2}\mu_{-}^{2}}_{q}\,.
\end{equation}

\ni If the electrons density in the absence of pair production is named $n_o$, $$n_- = n_o + n_+\,\,,$$ which provides 
$$n_+ = - \frac n2\,+\, \Big[\frac{n_o^2}{4}\,+\,n_+\,n_- \Big]^{1/2} \,\,.$$

Next section we will investigate analogous reactions using another non-Gaussian thermostatistical formalism developed by Kaniadakis, namely, the $\kappa$-statistics, as we will see just below.   Although there is relation between both parameters, $q$ and $\kappa$, we will find some particular features due only to Kaniadakis formulation.

\section{Kaniadakis statistics and the Saha equation}

The well known Kaniadakis statistics  \cite{Kaniadakis01a}, also well known as the $\kappa$-statistics, analogously to Tsallis thermostatistics model, generalizes the usual BG statistics initially by introducing both the $\kappa$-exponential and $\kappa$-logarithm defined respectively by
\begin{eqnarray}
\label{expk}
e_\kappa^{(f)}\,=\,
\Big( \sqrt{1+\kappa^2 f^2}+\kappa f \Big)^{1/\kappa}
\end{eqnarray}
and
\begin{eqnarray}
\label{logk}
\ln_\kappa(f)=\frac{f^\kappa-f^{-\kappa}}{2\kappa}\,\,.
\end{eqnarray}

\ni The following property can be satisfied, namely,
\begin{eqnarray}
\ln_\kappa\Big(\exp_\kappa(f)\Big)=\exp_\kappa\Big(\ln_\kappa(f)\Big)\equiv f\,\,.
\end{eqnarray}

\ni From Eqs. (\ref{expk}) and (\ref{logk}) we can notice that the $\kappa$-parameter twists the exponential and logarithm functions standard definitions.

The $\kappa$-entropy, connected to this $\kappa$-framework, can be written as
\begin{eqnarray}
S_\kappa=- k_B \sum_i^W  \,\frac{p_i^{1+\kappa}-p_i^{1-\kappa}}{2\kappa}\,\,,
\end{eqnarray}

\ni which recovers the BG entropy in the limit $\kappa \rightarrow 0$. It is important to say here that the $\kappa$-entropy satisfies the properties concerning concavity, additivity and extensivity. The $\kappa$-statistics has thrived when applied in many experimental scenarios. As an example we can cite cosmic rays  \cite{Kanisca1} and cosmic effects  \cite{aabn-1}, quark-gluon plasma  \cite{Tewe}, kinetic models describing a gas of interacting atoms and photons  \cite{Ross} and financial models  \cite{RBJ}.

%\subsection{Kaniadakis path}

Following the same procedure adopted before, concerning the Tsallis method, but now using the Kaniadakis $\kappa$-statistics, we assume that the chemical potential is related to the number density and temperature by a partition function

\begin{equation}
    z=e^{\beta\mu}_{\kappa}\,,
\end{equation}

\noindent where $e^{f}_{\kappa}$ was defined in Eq. (\ref{expk}). Then, extending  Eq. (\ref{cp}) in order to introduce the Kaniadakis statistics, we have that

%\noindent %where \( e^{f}_{\kappa}\equiv \left(\sqrt{1+\kappa^{2}f^{2}}+\kappa f\right)^{1/\kappa}\).  From Eq.(\ref{cp}) we have

%\begin{equation}
    %z=\frac{n\hbar^{3}}{g_{s}}\left(\frac{2\pi}{m K_{B}T}\right)^{3/2}\,.
%\end{equation}

%so

\begin{equation}\label{emuk}
    e^{\beta\mu}_{\kappa}=\frac{n\hbar^{3}}{g_{s}}\left(\frac{2\pi}{m k_{B}T}\right)^{3/2}\,.
\end{equation}

%%%%%%%%%%%%%%%%%%%%%%%%%
%For a given $i$ particle, equation (\ref{emuk}) becomes \(\beta \mu_i = \ln_\kappa \left( \frac{n_i}{g_{s_{i}}} \left( \frac{2 \pi \hbar^2}{m_i K_B T} \right)^{3 / 2}\right) \) where $\ln_\kappa %f\equiv\frac{f^{\kappa}-f^{-\kappa}}{2\kappa}$. As $n_{i}=yn$
%\begin{equation}\label{lki}
%\beta\mu_i=\ln_\kappa\left(\frac{y}{2B}\frac{\rho}{T^{3/2}}\right)
%\end{equation}
%where $B=C$ for electrons or $B=C\alpha^{3/2}$ for protons. 
%%%%%%%%%%%%%%%%%%%%%%%%

\noindent Following the same procedure used in the last section, we have the ionization populations defined with the Kaniadakis formalism such as

\begin{equation}
n_{e}=\frac{g_{s_{e}}}{\hbar^{3}}\left(\frac{m_{e} k_{B}T}{2\pi}\right)^{3/2}e^{\beta\mu_{e}}_{\kappa}\,,
\end{equation}

\begin{equation}
n_{p}=\frac{g_{s_{p}}}{\hbar^{3}}\left(\frac{m_{p} k_{B}T}{2\pi}\right)^{3/2}e^{\beta\mu_{p}}_{\kappa}\,,
\end{equation}
and
\begin{equation}
n_{H}=\frac{g_{s_{H}}}
{\hbar^{3}}\left(\frac{m_{H} k_{B}T}{2\pi}\right)^{3/2}e^{\beta(\mu_{H}+|\epsilon_{o}|)}_{\kappa}\,,
\end{equation}
which implies that
\begin{equation}
\qquad \frac{n_{e}n_{p}}{n_{H}}=\frac{g_{s_{e}}g_{s_{p}}}{g_{s_{H}}}\left(\frac{m_{e}m_{p}k_{B}T}{(m_{p}+m_{e})2\pi\hbar^{2}}
\right)^{3/2}\frac{e^{\beta\mu_{e}}_{\kappa}e^{\beta\mu_{p}}_{\kappa}}{e^{\beta(\mu_{H}+|\epsilon_{o}|)}_{\kappa}}\,\,.
\end{equation}

\noindent After the usual simplification we obtain:

\begin{equation}
\frac{n_{e}^{2}}{n_{H}}=\left(\frac{m_{e}k_{B}T}{2\pi\hbar^{2}}\right)^{3/2}\frac{e^{\beta\mu_{e}}_{\kappa}e^{\beta\mu_{p}}_{\kappa}}
{e^{\beta(\mu_{H}+|\epsilon_{o}|)}_{\kappa}}\,.
\end{equation}

\noindent Using the definition of ionization fraction we obtain
 
\begin{equation}
\frac{y^{2}}{1-y}= \frac{1}{n}\left(\frac{m_{e}k_{B}T}{2\pi\hbar^{2}}\right)^{3/2}
\frac{e^{\beta\mu_{e}}_{\kappa}e^{\beta\mu_{p}}_{\kappa}}{e^{\beta(\mu_{H}+|\epsilon_{o}|)}_{\kappa}}\,,
\end{equation}

\noindent and using the $\kappa$-algebraic properties \cite{Kaniadakis01a} 

\begin{equation}
e^{x}_{\kappa}\,e^{y}_{\kappa}=e^{(x\,\overset{\kappa}{\oplus}\,y)}_{\kappa}=e^{\left(x\sqrt{1+\kappa^{2}y^{2}}+y\sqrt{1
+\kappa^{2}x^{2}}\right)}_{\kappa}\,,
\end{equation}

%\begin{equation}
%\frac{e^{x}_{\kappa}}{e^{y}_{\kappa}}=e^{x}_{\kappa}\,e^{-y}%_{\kappa} \\
%\textrm{\bf verificar!!}
%\end{equation}

\noindent we have

\begin{equation}
\frac{e^{\beta\mu_{e}}_{\kappa}e^{\beta\mu_{p}}_{\kappa}}{e^{\beta(\mu_{H}+|\epsilon_{o}|)}_{\kappa}}=
\frac{\exp_{\kappa} \left(\beta\mu_{e}\sqrt{1+\kappa^{2}\beta^{2}\mu_{p}^{2}}+\beta\mu_{p}\sqrt{1+\kappa^{2}\beta^{2}\mu_{e}^{2}}\right)}
{e_\kappa^{(\beta\mu_{H}+\beta|\epsilon_{o}|)}} \,.
\end{equation}

\noindent Finally, the ionization fraction is determined by 

\begin{equation} %\label{kalge1}%
\frac{y^{2}}{1-y}= \frac{1}{n}\left(\frac{m_{e}k_{B}T}{2\pi\hbar^{2}}\right)^{3/2}
e^{(\Theta_{\kappa 1} - \Theta_{\kappa 2})}_\kappa,
\end{equation}
where $\Theta_{\kappa1}$ is defined by

\begin{equation}
\Theta_{\kappa1}\equiv\left(\beta\mu_{e}\sqrt{1+\kappa^{2}\beta^{2}\mu_{p}^{2}}+\beta\mu_{p}\sqrt{1+\kappa^{2}
\beta^{2}\mu_{e}^{2}}\right)\sqrt{1+\kappa^{2}(\beta\mu_H+\beta|\epsilon_{o}|)^{2}}\,,
\end{equation}

\noindent and $\Theta_{\kappa2}$ is given by

\begin{equation}
\Theta_{\kappa2}\equiv(\beta\mu_H+\beta|\epsilon_{o}|)\sqrt{1+\kappa^{2}\left(\beta\mu_{e}
	\sqrt{1+\kappa^{2}\beta^{2}\mu_{p}^{2}}+\beta\mu_{p}\sqrt{1+\kappa^{2}\beta^{2}\mu_{e}^{2}}\right)^{2}}\,.
\end{equation}

%with $\beta\mu_i$ being defined in (\ref{lki}).

\noindent Also, as carried out before, for pure hydrogen we have an equation for the ionization fraction given by

\begin{equation} 
\frac{y^{2}}{1-y}= \frac{4.01\times 10^{-9}}{\rho}\,\,T^{\;3/2}\; e^{(\Theta_{\kappa1}-\Theta_{\kappa2})}_\kappa\,\,,
\end{equation}
where $\rho$ is the density in grams per cubic centimeter and $T$ is the Kelvin temperature.

\section{Conclusions}

The Saha-Langmuir equation, commonly known as Saha equation, provides the statistical distribution between ionization levels.  It can alternatively use the partition function for both states.   One of its main application is relative to stellar statistical distribution and plasma physics.   It is analogous to the Maxwell-Boltzmann velocity distribution statistics.   So, it is natural to explore alternative statistical formulations to search for the best method to analyze a particular physical system.   This is exactly the objective of this work, where we have used two different non-Gaussian statistical framework that are very successful in several areas of research inside and outside theoretical and experimental physics.

Through the point of view of Tsallis and Kaniadakis thermostatistical formalisms, we have constructed the Saha relations for some ionization reactions such as proton/electron, complex atoms and pair production.   We have constructed a chemical $q$-potential for each one of these ionization reactions and the ionization fractions were computed.

We have analyzed the asymptotic behavior of the chemical $q$-potential as a function of the temperature for several $q$-values through their curves.   The graphic has shown clearly, besides the asymptotic behavior, that the chemical $q$-potential is very sensible to the sign of the $q$-parameter.   We have demonstrated that the curves are totally symmetric.   We obtained a $q$-plus and a $q$-minus regions, and the boundary between both is a straight line for the $q=1$ value, i.e., it is the point where the BG statistical approach is recovered.   It is a very interesting and new result.   Since the relation between both statistics is $k=q-1$, the shape of the same curves in Kaniadakis procedure is analogous.

\acknowledgments

\ni E.M.C.A.  and J.A.N. thank CNPq (Conselho Nacional de Desenvolvimento Cient\' ifico e Tecnol\'ogico), Brazilian scientific support federal agency, for partial financial support, Grants numbers 406894/2018-3 and 302155/2015-5 (E.M.C.A.), and 303140/2017-8 (J.A.N.).

\end{document}